\documentstyle[12pt]{article}
\begin{document}

\def\ds{\displaystyle}
\def\beq{\begin{equation}}
\def\eeq{\end{equation}}
\def\bea{\begin{eqnarray}}
\def\eea{\end{eqnarray}}
\def\beeq{\begin{eqnarray}}
\def\eeeq{\end{eqnarray}}
\def\ve{\vert}
\def\vel{\left|}
\def\ver{\right|}
\def\nnb{\nonumber}
\def\ga{\left(}
\def\dr{\right)}
\def\aga{\left\{}
\def\adr{\right\}}
\def\lla{\left<}
\def\rra{\right>}
\def\rar{\rightarrow}
\def\nnb{\nonumber}
\def\la{\langle}
\def\ra{\rangle}
\def\ba{\begin{array}}
\def\ea{\end{array}}
\def\tr{\mbox{Tr}}
\def\ssp{{\Sigma^{*+}}}
\def\sso{{\Sigma^{*0}}}
\def\ssm{{\Sigma^{*-}}}
\def\xis0{{\Xi^{*0}}}
\def\xism{{\Xi^{*-}}}
\def\qs{\la \bar s s \ra}
\def\qu{\la \bar u u \ra}
\def\qd{\la \bar d d \ra}
\def\qq{\la \bar q q \ra}
\def\gGgG{\la g^2 G^2 \ra}
\def\q{\gamma_5 \not\!q}
\def\x{\gamma_5 \not\!x}
\def\g5{\gamma_5}
\def\sb{S_Q^{cf}}
\def\sd{S_d^{be}}
\def\su{S_u^{ad}}
\def\ss{S_s^{??}}
\def\sbp{{S}_Q^{'cf}}
\def\sdp{{S}_d^{'be}}
\def\sup{{S}_u^{'ad}}
\def\ssp{{S}_s^{'??}}
\def\sig{\sigma_{\mu \nu} \gamma_5 p^\mu q^\nu}
\def\fo{f_0(\frac{s_0}{M^2})}
\def\ffi{f_1(\frac{s_0}{M^2})}
\def\fii{f_2(\frac{s_0}{M^2})}
\def\O{{\cal O}}
\def\sl{{\Sigma^0 \Lambda}}
\def\es{\!\!\! &=& \!\!\!}
\def\ar{&+& \!\!\!}
\def\ek{&-& \!\!\!}
\def\cp{&\times& \!\!\!}
\def\se{\!\!\! &\simeq& \!\!\!}
\def\kpm{&\pm& \!\!\!}
\def\kmp{&\mp& \!\!\!}
\def\arr{\!\!\!\!&+&\!\!\!}
\def\eqv{&\equiv& \!\!\!}

\def\simlt{\stackrel{<}{{}_\sim}}
\def\simgt{\stackrel{>}{{}_\sim}}


\title{
         {\Large
                 {\bf
$\phi \rar KK$ decay in light cone QCD
                 }
         }
      }

\author{\vspace{1cm}\\
{\small T. M. Aliev \thanks
{e-mail: taliev@metu.edu.tr}\,\,,
\.{I} Kan{\i}k \thanks
{e-mail: e114288@metu.edu.tr}\,\,,
M. Savc{\i} \thanks
{e-mail: savci@metu.edu.tr}} \\
{\small Physics Department, Middle East Technical University, 
06531 Ankara, Turkey} }

\date{}

\begin{titlepage}
\maketitle
\thispagestyle{empty}

\begin{abstract}
   
The coupling constant of $\phi \rar KK$ decay is calculated in light cone   
QCD sum rules. The result obtained for $g_{\phi KK} = (4.9 \pm 0.8)$ is in a
good agreement with the existing experimental result.

\end{abstract}

~~~PACS numbers: 11.50.Hx, 13.25.Jx
\end{titlepage}

\section{Introduction}
Light scalar mesons constitute a remarkable exception of the quark model
systematization of mesons and their nature still need to be unambiguously
established \cite{R5601}. 

Particularly, the nature $f_0(980)$ meson is under debate. According to the
naive $\bar{q}q$ picture and strong coupling with kaons, $f_0(980)$ can be
interpreted as a pure $\bar{s}s$ state \cite{R5602}--\cite{R5604}. However, this
interpretation does not explain mass degeneracy between $f_0(980)$ and
isovector $a_0(980)$, which is interpreted as a
($\bar{u}u-\bar{d}d)/\sqrt{2}$ state. It is also interpreted as a four quark
$\bar{q}q\bar{s}s$ \cite{R5605} bound state of hadrons
\cite{R5606}--\cite{R5608} and as a result of a process known as hadronic
dressing \cite{R5602,R5609}. 

For understanding the content of the $f_0$ meson several alternatives have 
been suggested: For example, analysis of $\phi \rar f_0 \gamma$ decay 
\cite{R5605}--\cite{R5610} and investigation of the ratio 
$\Gamma(a_0 \rar f_0 \gamma)/\Gamma(\phi \rar f_0 \gamma)$ \cite{R5607,R5608} 
are believed  to be the most promising ones for this purpose. 

The $\phi \rar f_0 \gamma$ decay is a very efficient tool for this purpose,
since the branching ratio is essentially dependent on the content of $f_0$.
For example, if $f_0$ is a pure $\bar{s}s$ state, the branching ratio is
$\sim 10^{-5}$, while if $f_0$ is composed of four quarks then the branching
ratio is expected to be $\sim 10^{-4}$. 

The strong coupling constants $g_{\phi K^+ K^-}$ and $g_{f_0 K^+ K^-}$ are 
among the important hadronic parameters entering to the analysis involving
$\phi$ and $f_0(980)$. Indeed, the kaon loop diagrams contributing 
$\phi \rar f_0 \gamma$ are expected to be in terms of $g_{f_0 K^+ K^-}$, as 
well as $g_{\phi K^+ K^-}$. The coupling constant $g_{f_0 K^+ K^-}$ is studied 
in light cone QCD sum rules \cite{R5610} (more about light cone QCD sum rules 
and and its applications can be found in \cite{R5611,R5612}).

In the present work we calculate the strong coupling constant 
$g_{\phi K^+ K^-}$ in light cone QCD sum rules method. It should be noted
that this constant can be obtained from experimental data on $\phi$ meson
decays. The goal in the present work is twofold: Firstly, can we get new
information about the quark content of $\phi$ meson comparing experimental
data with theoretical results? Secondly, how does light cone QCD work for
the asymmetric case, i.e., with different Borel mass parameters corresponding 
to different mass channels?  

The paper is organized as follows. In section 2, we derive sum rules for the 
$g_{\phi K^+ K^-}$ coupling constant. In section 3, we present our numerical
results and conclusion.

\section{Sum rules for $g_{\phi K^+ K^-}$ coupling constant}

In this section we calculate the strong coupling constant $g_{\phi K^+ K^-}$
in light cone QCD sum rules. This coupling constant is defined by the
following matrix element:
\bea
\label{e5601}
\la K^-(q) \phi^0(p,\varepsilon) \ve K^+(p+q) \ra~,
\eea 
where the momentum assignment is specified in brackets and $\varepsilon_\mu$
is the polarization vector of the $\phi$ meson. In order to calculate the
strong  coupling constant $g_{\phi K^+ K^-}$ we consider the following
correlator function
\bea
\label{e5602}
\Pi_{\mu\nu}(p,q) = i \int d^4x e^{ipx} \lla K(q) \vel \mbox{\rm T} \left\{
J_\nu^\phi(x) \bar{J}_\mu^K (0) \right\}\ver 0 \rra~,
\eea
where the quark current $J_\mu^K = \bar{u} \gamma_\mu \gamma_5 s$ is the
axial vector current and $J_\nu^\phi = \bar{s} \gamma_\mu s$ is the
interpolating current for the $\phi$ meson.

The correlator function, in general, can be written in terms of the
following five independent invariant functions
\bea
\label{e5603}
\Pi_{\mu\nu}(p,q) = \Pi_1 g_{\mu\nu} + \Pi_2 p_\mu p_\nu + \Pi_3 p_\mu q_\nu
+ \Pi_4 q_\mu p_\nu + \Pi_5 q_\mu q_\nu~.
\eea
Therefore, our first problem is to choose the kinematical structure. For
this aim, we consider the phenomenological part of the correlator function.
This part can be written as
\bea
\label{e5604}
\Pi_{\mu\nu} = \sum \frac{\la K^-(q) \phi^0(p) \ve K^+(p+q) \ra 
\la K^+(p+q) \ve J_\mu^K \ve 0 \ra
\la 0 \ve J_\nu^\phi \ve \phi(p)\ra}
{(p^2-m_\phi^2) [(p+q)^2-m_K^2]}~.
\eea
The matrix elements entering Eq. (\ref{e5604}) are defined as
\bea
\label{e5605}
\la K^+(p+q) \ve J_\mu^K \ve 0 \ra \es f_K (p+q)_\mu~, \nnb \\
\la 0 \ve J_\nu^\phi \ve \phi(p)\ra \es m_\phi f_\phi \varepsilon_\nu~.
\eea
Using Eqs. (\ref{e5604}) and (\ref{e5605}), we get for the
physical part
\bea
\label{e5606}
\Pi_{\mu\nu} = \frac{g_{\phi K^+ K^-} f_K m_\phi f_\phi}
{(p^2-m_\phi^2) [(p+q)^2-m_K^2]}\, (p_\mu+q_\mu) \ga q_\nu + \frac{1}{2} p_\nu
\dr~.
\eea
It follows from this expression that the only the structures $p_\mu q_\nu$,
$q_\mu q_\nu$, $q_\mu p_\nu$ and $p_\mu p_\nu$ give contribution to the
correlator function. In further analysis, we will choose the structure
$p_\mu q_\nu$ from which the corresponding invariant structure
\bea
\label{e5607}
\Pi = \frac{g_{\phi K^+ K^-} f_K m_\phi f_\phi}{(p^2-m_\phi^2)
[(p+q)^2-m_K^2]}~,
\eea
follows.

Our next task is the calculation of the correlator function from QCD side.
This calculation can be carried out by using light cone operator product 
expansion method, in which we work with large momenta, i.e., $-p^2$ and
$-(p+q)^2$ are both large. The correlator function, then, can be calculated
as an expansion near to the light cone $x^2 \simeq 0$. The expansion
involves matrix elements of the nonlocal operators between vacuum and the
kaon states, i.e., in terms of kaon wave functions with increasing twist.

After lengthy calculations, we get the following expression for the
invariant function which is proportional to the structure $p_\mu q_\nu$ 
\bea
\label{e5608}
\lefteqn{
\Pi(p^2,(p+q)^2) = i f_K \int_0^1 du \Bigg\{
\frac{4 u g_2(u)}{\Delta^2} +\frac{\varphi_K(u)}{\Delta}
- 4 \frac{g_1(u) + G_2(u)}{\Delta^2} \ga 1+ \frac{2 m_s^2}{\Delta}\dr} \nnb \\
\ar \frac{m_K^2}{3} \frac{\varphi_\sigma(u)}{\Delta^2}\Bigg\} +
i f_K \int_0^1 du\, u \int {\cal D} \alpha_i \frac{2}{\Delta_2^2} 
\Big[2 \varphi_\perp(\alpha_i) + \varphi_\parallel + 
2\tilde{\varphi}_\perp(\alpha_i) \Big] \nnb \\
\ar if_K \int_0^1 du \int_0^1 d\alpha_3 \int_0^{1-\alpha_3} d \alpha_1
\frac{1}{\Delta_2^2} \Big[ 2 \varphi_\perp(\alpha_i) - 
\varphi_\parallel + 2 \tilde{\varphi}_\perp(\alpha_i)- 
\tilde{\varphi}_\parallel(\alpha_i) \Big] \nnb \\
\ar 2 i f_K \Bigg\{\int_0^1 du(u-1) \int_0^1 d\alpha_3 \frac{4 \hat{F}(\alpha_3) 
\{ pq + m_K^2 [1 + \alpha_3(u-1) ] \}}{\Delta_1^3} \nnb \\
\ar \int_0^1 du \int_0^1 d\alpha_3 \int_0^{1-\alpha_3} d\alpha_1
\frac{4 F(\alpha_i) [ pq + m_K^2 (\alpha_1 + u \alpha_3) ]}
{\Delta_2^3}~\Bigg\},
\eea
where 
\bea
\label{e5609}
\Delta \es m_s^2 -(p+q u)^2~, \nnb \\
\Delta_1 \es m_s^2 - [p+q (1+(u-1) \alpha_3]^2~, \nnb \\
\Delta_2 \es m_s^2-[p+q (\alpha_1+u \alpha_3)]^2~,
\eea
and,
\bea
\label{e5610}
\hat{F}(\alpha_3) \es -\int_0^{\alpha_3} dt \int_0^{1-t} d \alpha_1
\Phi(\alpha_1,1-\alpha_1-t,t)~, \\ \nnb \\
\label{e5611}
F(\alpha_i) \es - \int_{0}^{\alpha_1} dt \Phi(t,1-\alpha_3-t,\alpha_3)~, 
\\ \nnb \\
\label{e5612}
\Phi(\alpha_i) \es \varphi_\parallel (\alpha_i) + \varphi_\perp (\alpha_i)
+\tilde{\varphi}_\parallel (\alpha_i) + \tilde{\varphi}_\perp (\alpha_i)~.   
\eea

The functions in Eq. (\ref{e5608}) are defined as
\bea
\label{e5613}
\la K(q) \ve \bar{u}(x) \gamma_\mu \gamma_5 s(0) \ve 0 \ra \es 
-i f_K q_\mu \int_0^1 du e^{iuqx} [\varphi_K(u) + x^2 g_1(u) ] \nnb \\
\ar f_K \ga x_\mu - \frac{q_\mu x^2}{qx}\dr 
\int_0^1 du e^{iuqx}g_2(u)~, \\ \nnb \\
\label{e5614}
\la K(q) \ve \bar{u}(x) \sigma_{\mu\nu} \gamma_5 s(0) \ve 0 \ra \es
i (q_\mu x_\nu - q_\nu x_\mu) \frac{f_K m_K^2}{6 m_s} 
\int_0^1 du e^{iuqx} \varphi_\sigma (u)~,
\eea
and
\bea
\label{e5615}
G(u) = - \int_0^u g_2(u) dv~.
\eea

The matrix elements involving quark--gluon field are determined as
\bea
\label{e5616}
\lefteqn{
\la K(q) \ve \bar{u}(x) \gamma_\mu \gamma_5 g_s
G_{\alpha\beta} (ux) s(0) \ve 0 \ra =} \nnb \\
&&f_K \Bigg[ q_\beta \ga g_{\alpha\mu} - \frac{x_\alpha q_\mu}{qx} \dr
- q_\beta \ga g_{\beta\mu} - \frac{x_\beta q_\mu}{qx} \dr \Bigg] 
\int {\cal D} \alpha_i \varphi_\perp (\alpha_i) e^{i qx(\alpha_1+ u
\alpha_3)} \nnb \\
\ar f_K \frac{q_\mu}{qx} \ga q_\alpha x_\beta - q_\beta x_\alpha \dr
\int {\cal D} \alpha_i \varphi_\parallel (\alpha_i) e^{i qx(\alpha_1+ u
\alpha_3)}\\ \nnb \\
\label{e5617}
\lefteqn{
\la K(q) \ve \bar{u}(x) \gamma_\mu g_s \tilde{G}_{\alpha\beta}(ux) 
s(0) \ve 0 \ra =} \nnb \\
&&i f_K \Bigg[ q_\beta \ga g_{\alpha\mu} - \frac{x_\alpha q_\mu}{qx} \dr
- q_\beta \ga g_{\beta\mu} - \frac{x_\beta q_\mu}{qx} \dr \Bigg]
\int {\cal D} \alpha_i \tilde{\varphi}_\perp (\alpha_i) e^{i qx(\alpha_1+ u
\alpha_3)} \nnb \\
\ar i f_K \frac{q_\mu}{qx} \ga q_\alpha x_\beta - q_\beta x_\alpha \dr
\int {\cal D} \alpha_i \tilde{\varphi}_\parallel (\alpha_i) e^{i qx(\alpha_1+ u
\alpha_3)}~,
\eea  
where $\tilde{G}_{\alpha\beta} = \frac{1}{2}
\epsilon_{\alpha\beta\rho\sigma} G^{\rho\sigma}$, ${\cal D}\alpha_i
=d\alpha_1 d\alpha_2 d\alpha_3 \delta(1-\alpha_1-\alpha_2-\alpha_3)$. 

The sum rule for $g_{\phi K^+ K^-}$ is obtained by equating the
phenomenological, Eq. (\ref{e5607}), and theoretical, Eq. (\ref{e5608}),
parts.

In order to suppress the contributions of the continuum and higher states,
we perform double Borel transformation over the variables $-p^2$ and
$-(p+q)^2$ on both sides of Eqs. (\ref{e5607}) and (\ref{e5608}), and 
obtain the following expression for the correlator function
\bea
\label{e5618}
\lefteqn{
f_K m_\phi f_\phi g_{\phi K K} e^{-m_phi^2/M_1^2} e^{-m_K^2/M_2^2} =
f_K e^{-m_0^2/M^2} \Bigg\{ M^2 \varphi_K(u_0) + 4 u_0 g_2(u_0)} \nnb \\
\ek 4 [g_1(u_0) + G_2(u_0)] + \frac{m_K^2}{3} \varphi_\sigma(u_0) - 
4 \frac{m_s^2}{M^2} [g_1(u_0) + G_2(u_0)]
+ \Bigg(\int_0^{1-u_0} d\alpha_3 
\int_{u_0-\alpha_3}^{u_0}d\alpha_1 \nnb \\
\ar \int_{1-u_0}^{1} d\alpha_3 \int_{u_0-\alpha_3}^{1-\alpha_3}d\alpha_1 -
\int_{u_0}^{1} d\alpha_3 \int_{u_0-\alpha_3}^{0}d\alpha_1 \Bigg)
\Bigg( 2 \frac{u_0-\alpha_1}{\alpha_3} \,
\frac{2 \varphi_\perp(\alpha_i)+\varphi_\parallel(\alpha_i)+
2 \tilde{\varphi}_\perp(\alpha_i)}{\alpha_3} \nnb \\
\ar \frac{\Phi (\alpha_i)}{\alpha_3} - \frac{2}{\alpha_3} 
\frac{dF(\alpha_i)}{d\alpha_1} \Bigg) 
- 2 \int_{1-u_0}^1 \frac{\hat{F}^\prime (\alpha_3)}{\alpha_3} d\alpha_3 - 2
 \int_{1-u_0}^1 d\alpha_3 \frac{F(1-\alpha_3,0,\alpha_3)}{\alpha_3}
\Bigg\}~,
\eea 
where
\bea
M^2 = \frac{M_1^2 M_2^2}{M_1^2 + M_2^2}~,~~~~~
u_0 = \frac{M_1^2}{M_1^2 + M_2^2}~,~~~~\mbox{and},~~~~~
m_0^2 = m_s^2 + m_K^2 u_0 (1-u_0)~. \nnb
\eea
Subtraction of the continuum and higher states is carried out by employing
the quark--hadron duality, i.e., continuum contribution, which is
represented in terms of the spectral density obtained from QCD side, by
equating it to the one obtained from QCD side, but starting from some given
threshold. The prescription for subtraction the contribution of the
continuum in light cone version of the sum rule is proposed in \cite{R5613}
(see also \cite{R5614}). In \cite{R5613} and in many works, the symmetric
point $M_1^2=M_2^2=2 M^2$ (i.e., $u_0=1/2$) is considered, and then the
continuum subtraction is implemented by means of the simple substitution
\bea
e^{-m^2/M^2} \rar e^{-m^2/M^2} - e^{-s_0/M^2}~, \nnb
\eea
in the leading twist term (in our case leading twist term is the wave
function $\varphi_K(u)$). But this prescription is not adequate in our case,
where the Borel parameters and masses of different channels are not equal.
In the present work we will follow the analysis given  in \cite{R5610},
where the prescription for continuum subtraction through use of the Borel 
parameters with different masses in the respective channels is proposed,
and properties of the wave functions are exploited. Namely, the leading
twist--2 wave function can be exploited as a power series 
\bea
\varphi_K(u) = \sum_k b_k (1-u)^k~, \nnb
\eea
in order to calculate its contribution in the duality region. Here we will
neglect the continuum subtraction in the higher twist terms altogether, due
to their small contribution to the theoretical part of the sum rules.
Here, we will neglect the continuum subtraction in all higher twist terms,
due to their small contribution to the theoretical part of the sum rules.

The final result for the $g_{\phi KK}$ coupling is given as
\bea
\label{e5619}
\lefteqn{
g_{\phi KK} = \frac{1}{m_\phi f_\phi} e^{m_\phi^2/M_1^2} e^{m_K^2/M_2^2}
e^{-m_0^2/M^2}
\Bigg\{ M^2 \sum_k b_k \ga \frac{M^2}{M_1^2} \dr^k 
\Bigg[ 1-e^{-(s_0-m_s^2)/M^2} \sum_{i=0}^k \frac{1}{i!} \ga
\frac{s_0-m_s^2}{M^2} \dr^i } \nnb \\ 
\ar e^{-(s_0-m_s^2)/M^2}\frac{m_K^2 M^2}
{M_1^2 M_2^2} \frac{1}{(k+1)!}\ga \frac{s_0-m_s^2}{M^2} \dr^{k+1} \Bigg]     
 + 4 u_0 g_2(u_0) - 4 [g_1(u_0) + G_2(u_0) ] \nnb \\
\ar \frac{m_K^2}{3} \varphi_\sigma
(u_0) - 4 \frac{m_s^2}{M^2} [g_1(u_0) + G_2(u_0) ] +  \Bigg(
\int_0^{1-u_0} d\alpha_3 \int_0^{u_0} d\alpha_1 +
\int_{1-u_0}^1 d\alpha_3 \int_{u_0-\alpha_3}^{1-\alpha_3} d\alpha_1 \nnb \\
\ek \int_{u_0}^1 d\alpha_3 \int_{u_0-\alpha_3}^0 d\alpha_1 \Bigg)
\Bigg[2 \frac{u_0-\alpha_1}{\alpha_3^2} \Big( 2 \varphi_\perp (\alpha_i) +
\varphi_\parallel (\alpha_i) + 2 \tilde{\varphi}_\perp (\alpha_i) \Big)
+ \frac{\Phi(\alpha_i)}{\alpha_3} - \frac{2}{\alpha_3} \frac{d F(\alpha_i)}
{d\alpha_1} \Bigg] \nnb \\
\ek 2 \int_{1-u_0}^1 {d\alpha_3} \frac{\hat{F}^\prime (\alpha_3)}
{\alpha_3} - 2 \int_{1-u_0}^1 {d\alpha_3}
\frac{F(1-\alpha_3,0,\alpha_3)}{\alpha_3} \Bigg\}~,
\eea
where $s_0$ is the smallest continuum contribution.

\section{Numerical analysis}

In this section we present our numerical calculation on $g_{\phi KK}$
coupling constant. It follows from Eq. (\ref{e5619}) that the main input
parameters are the kaon wave functions. The theoretical framework for their
determination is based on an expansion in terms of the matrix elements of
conformal operators \cite{R5615}. In particular, for the leading twist--2
wave function $\varphi_K(u)$ defined in Eq. (\ref{e5613}), the expansion 
goes into Gegenbauer polynomials:
\bea
\label{e5620}
\varphi_K (u,\mu^2) = 6u (1-u) \Bigg[ 1 +\sum_{n=1}^\infty a_{2n}(\mu^2)
C_{2n}^{3/2} (2u-1) \Bigg]~,
\eea
where $a_2(1~GeV)=0.2$ \cite{R5616}.

Analogously $\varphi_\sigma$ is defined as
\bea
\label{e5621}
\varphi_\sigma (u) = 6u (1-u)\Bigg[ 1 + \ga 5 \eta_3 - \frac{1}{2} \eta_3
w_3 - \frac{7}{20} \rho^2 - \frac{3}{5} \rho^2 \tilde{a}_2 \dr C_2^{3/2}
(2u-1) + \cdots \Bigg]~,
\eea
where, at the $\mu=1~GeV$ scale $\eta_3 = 0.015$, $w_3=-3$,
$\tilde{a}_2=0.2$. Here, the factor $\rho=m_s^2/m_K^2$ takes into account
the boson mass corrections (see \cite{R5617}). The twist--4 wave functions
$\varphi_\parallel (\alpha_i)$, $\varphi_\perp (\alpha_i)$,
$\tilde{\varphi}_\parallel (\alpha_i)$ and $\tilde{\varphi}_\perp
(\alpha_i)$, including the meson mass corrections are given as
(see \cite{R5615} and \cite{R5617})
\bea
\varphi_\perp (\alpha_i) \es 30 m_K^2 \alpha_3^2 (2 \alpha_1 -1 - \alpha_3) 
\Bigg[ h_{00} + h_{01} \alpha_3 + \frac{h_{10}}{2} (5\alpha_3-3) +\cdots
\Bigg] \nnb \\
\varphi_\perp (\alpha_i) \es 120 m_K^2 \alpha_1 (1 - \alpha_1 -\alpha_3)
\alpha_3 [a_{10} (1-2 \alpha_1 -\alpha_3)+\cdots \Bigg]~, \nnb \\
\tilde{\varphi}_\perp (\alpha_i) \es - 30  m_K^2 \alpha_3^2 
\Bigg\{ h_{00} (1-\alpha_3) + h_{01} [\alpha_3 (1-\alpha_3) +
6 \alpha_1 (1 - \alpha_1 -\alpha_3) \nnb \\ 
\ar h_{10} [\alpha_3 (1-\alpha_3)-
\frac{3}{2} [ \alpha_1^2 + (1 - \alpha_1 -\alpha_3)^2] +\cdots \Bigg\}~,\nnb \\
\tilde{\varphi}_\perp (\alpha_i) \es 120 m_K^2 \alpha_1 
(1 - \alpha_1 -\alpha_3)
\alpha_3 [v_{00} + v_{10}(3 \alpha_3-1)+\cdots \Bigg]~,\nnb
\eea
where
\bea
h_{00} \es v_{00} = -\frac{1}{3} \eta_4~,\nnb \\
h_{01} \es \frac{7}{4} \eta_4 w_4 - \frac{3}{20} a_2~, \nnb \\
h_{10} \es \frac{7}{2} \eta_4 w_4 + \frac{3}{20} a_2~, \nnb \\
v_{10} \es \frac{21}{8} \eta_4 w_4~, \nnb \\
a_{10} \es \frac{21}{8} \eta_4 w_4 - \frac{9}{20} a_2~, \nnb
\eea
with $\eta_4(\mu=1~GeV) = 0.6$ and $w_4(\mu=1~GeV) = 0.2$
\cite{R5615,R5617}.

The values of other input parameters appearing in Eq. (\ref{e5619}) are:
$m_s = 0.14~GeV$ \cite{R5618}, $m_K = 0.4937~GeV$, $m_\phi = 1.02~GeV$.
Leptonic decay constant of $\phi$ meson, $f_\phi= 0.234~GeV$, follows from
the experimental result of the $\phi \rar \ell^+ \ell^-$ decay \cite{R5619}.
The threshold $s_0$ which is varied around the value $s_0=1.1~GeV^2$, is
determined from the analysis of two--point function sum rules for $f_K$
\cite{R5620}. 

Having all input parameters, we now proceed by carrying out numerical
calculation. The dependence of $g_{\phi KK}$ on Borel masses $M_1^2$ and
$M_2^2$ at two fixed values of $s_0 = 1.1~GeV^2$ and $s_0 = 1.2~GeV^2$ is
presented in Figs. (1) and (2), respectively. According to the QCD sum rule
method ranges of the auxiliary Borel parameters $M_i^2$ should be found such
that  the result for $g_{\phi KK}$ be practically independent of them.

From these figures we see that, such regions indeed do exist. When $M_1^2$
and $M_2^2$ are varied in the regions $2~GeV^2 \le M_1^2 \le 4~GeV^2$ and 
$0.8~GeV^2 \le M_2^2 \le 1.4~GeV^2$, the result for $g_{\phi KK}$ seems to
be independent of the Borel parameters. It should be noted here that, the
result changes slightly when the continuum threshold is fixed to the value
$s_0=1.2~GeV^2$. The final result for $g_{\phi KK}$ is
\bea
\label{e5622}
g_{\phi KK} = 4.9 \pm 0.8~.
\eea

At this point, let us discuss sources of the uncertainties. $SU_f(3)$ 
breaking effects in kaon distribution amplitudes which we neglected, can 
play essential role, since we can explore wide range of $u$ and hence 
smoothing the effects of the shape of wave function. Additional uncertainty 
arises from the value of $m_s$. All these factors can cause an uncertainty 
about 5--10\%. Moreover, the errors coming from the variations in the
continuum threshold and Borel masses, change the result about 10\%. If all
these uncertainties are taken into account, the resulting error is about
20\%, which is quoted in Eq. (\ref{e5622}). 

Finally, we would comment that, existing experimental results on $\phi \rar
K K$ decay predicts $g_{\phi KK} = 4.8$. So, obviously, we see that our
result is quite close to the experimental value.  Therefore we conclude
that the quark content of $\phi$ is $\bar{s}s$, and for channels with
different masses and different Borel parameters, light cone QCD sum rules
work quite well.

\newpage

\section*{Figure captions}
{\bf Fig. (1)} The dependence of the coupling constant $g_{\phi KK}$  on 
the Borel parameters $M_1^2$ and $M_2^2$, at the fixed value $s_0=1.1~GeV^2$ 
of the continuum threshold.\\\\
{\bf Fig. (2)} The same as Fig. (1), but at the fixed value
$s_0=1.2~GeV^2$ of the continuum threshold.

\newpage

\begin{figure}  
\vskip 1.5 cm   
    \includegraphics{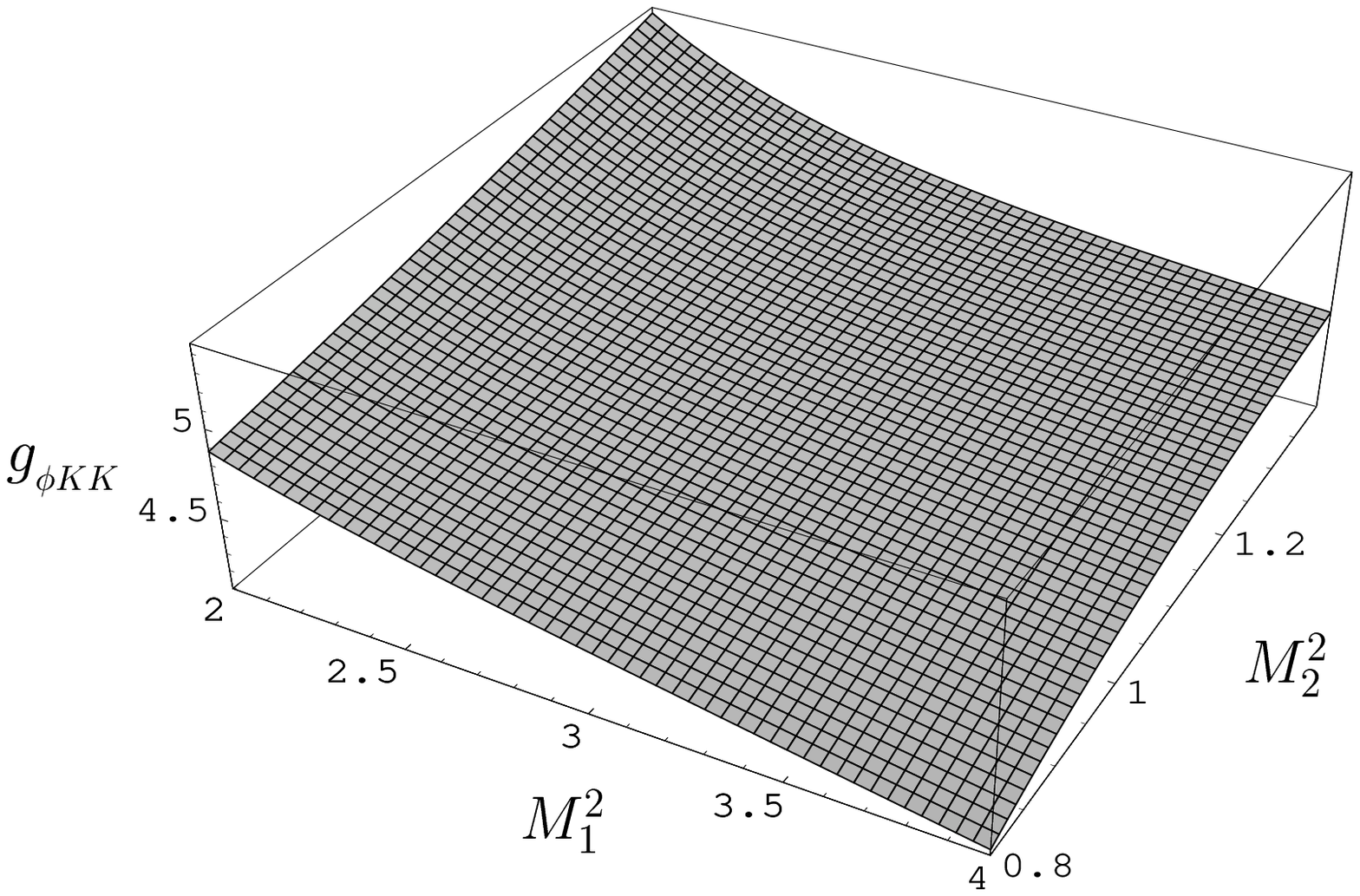}
\vskip 7.0cm     
\caption{}
\end{figure}

\begin{figure}  
\vskip 1. cm
    \includegraphics{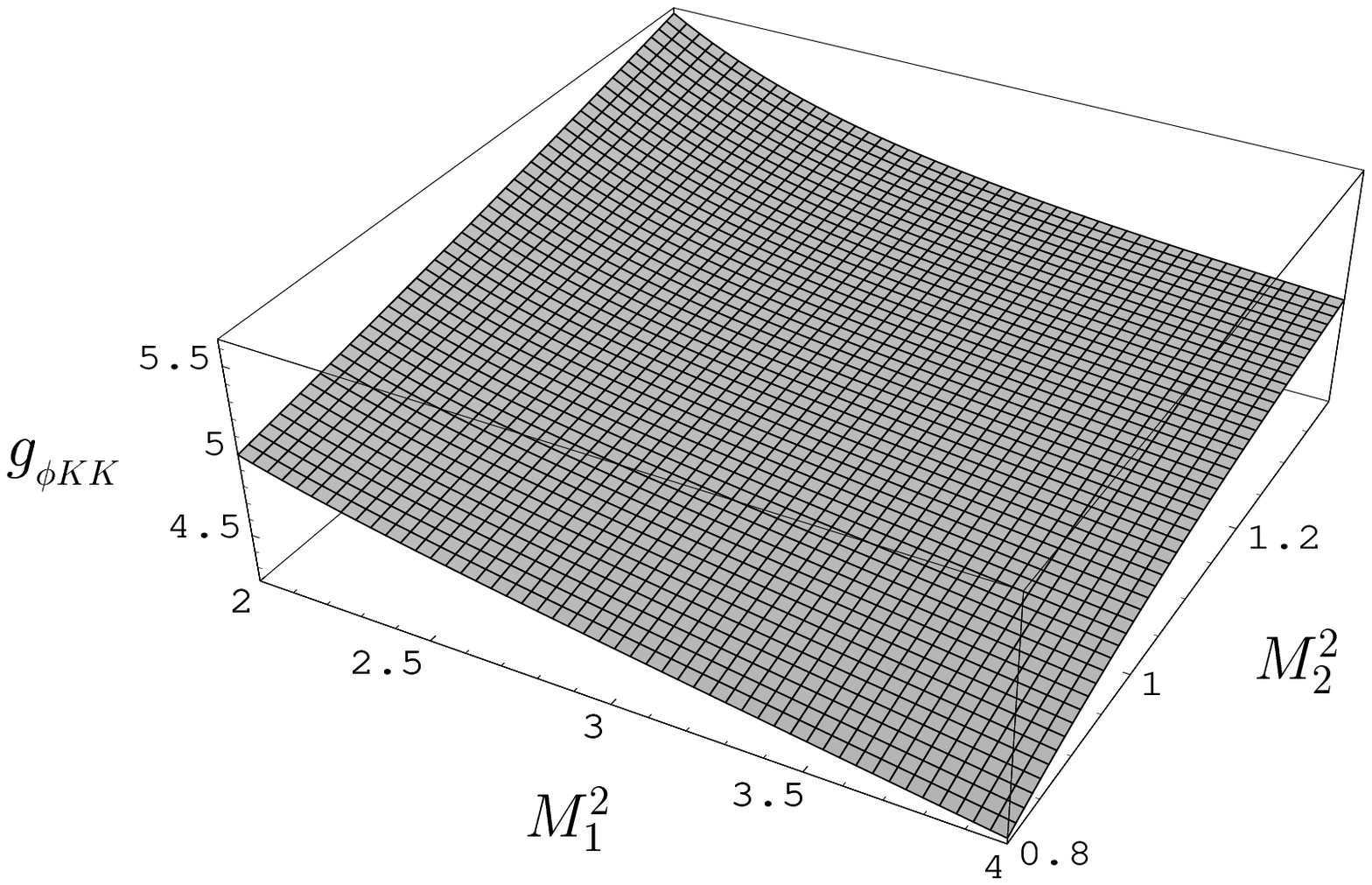}
\vskip 8.0 cm
\caption{}
\end{figure}

\end{document}